\documentclass[journal,comsoc]{IEEEtran}
\IEEEoverridecommandlockouts
\usepackage{amsmath,amssymb,amsfonts}
\usepackage{algorithmic}
\usepackage{graphicx,subfigure}
\usepackage{textcomp}
\usepackage{xcolor}
\usepackage{acronym} 
\usepackage{algorithm2e}
\usepackage[utf8]{inputenc}
\usepackage[backend=biber,style=ieee,defernumbers=true]{biblatex}

\newcommand\fig[1]{Figure~\ref{fig:#1}}

\newcommand\sect[1]{Section~\ref{sec:#1}}

\graphicspath{{./figures/}}

\newcommand{\figeps}[3][]{%
 \begin{figure}[htb]
  \begin{center}
   \leavevmode
      \parbox[t]{#1}{%
        \resizebox{#1}{!}{\includegraphics{#2}}
      }
      \vspace{-0.2cm}
   \caption{#3}
   \label{fig:#2}
  \end{center}
 \end{figure}
}

\newenvironment{tab}[2]{%
 \begin{table}[tbh]
  \begin{center}
  \caption{#2\label{tab:#1}}
}{%
  \end{center}
 \end{table}
}

\RestyleAlgo{ruled}
\begin{document}
\title{Impact of Packet Loss and Timing Errors on Scheduled Periodic Traffic with Time-Aware Shaping (TAS) in Time-Sensitive Networking (TSN)}
\author{
  \IEEEauthorblockN{Manuel Eppler, Steffen Lindner, Lukas Osswald, Thomas Stüber, Michael Menth}\\
  \IEEEauthorblockA{
  Chair of Communication Networks, University of Tuebingen, Germany\\
  \{manuel.eppler, steffen.lindner, lukas.osswald, thomas.stueber, menth\}@uni-tuebingen.de
  }}
\maketitle

\begin{abstract}
Time-Sensitive Networking (TSN) is a collection of mechanisms to enhance the realtime transmission capability of Ethernet networks. TSN combines priority queuing, traffic scheduling, and the Time-Aware Shaper (TAS) to carry periodic traffic with ultra-low latency and jitter. That is, so-called Talkers send periodic traffic with highest priority according to a schedule. The schedule is designed such that the scheduled traffic is forwarded by the TSN bridges with no or only little queuing delay. To protect that traffic against other frames, the TAS is configured on all interfaces such that lower-priority queues can send only when high-priority traffic is not supposed to be forwarded. In the literature on scheduling algorithms for the TAS there is mostly the explicit or implicit assumption that the TAS also limits transmission slots of high-priority traffic.

In this paper we show that this assumption can lead to tremendous problems like very long queuing delay or even packet loss in case of faulty frames. A faulty frame arrives too early or too late according to the schedule, it is missing or additional. We construct minimal examples to illustrate basic effects of faulty frames on a single link and demonstrate how this effect can propagate through the networks and cause remote problems. We further show using simulations that a single slightly delayed frame may lead to frame loss on multiple links. We show that these problems can be alleviated or avoided when TAS-based transmission slots for high-priority traffic are configured longer than needed or if they are not limited at all.  
\end{abstract}

\begin{IEEEkeywords}
TSN, Reliability, TAS, PSFP, OMNET++, Scheduled Traffic
\end{IEEEkeywords}

\subsection{List of frequently used acronyms}
We use the following acronyms.
\begin{acronym}[GRASP]
\acro{ATS}{Asynchronous Traffic Shaping}
\acro{BE}{Best Effort}
\acro{CBS}{Credit-Based Shaper}
\acro{CNC}{Centralized Network Configuration}
\acro{GCL}{Gate Control List}
\acro{gPTP}{generalized Precision Time Protocol}
\acro{IPV}{Internal Priority Value}
\acro{QoS}{Quality of Service}
\acro{TAS}{Time-Aware Shaper}
\acro{TSN}{Time-Sensitive Networking}
\acro{PSST}{Per-Stream Scheduled Traffic}
\acro{PCP}{Priority Code Point}
\acro{PSFP}{Per-Stream Filtering and Policing}
\acro{EST}{Enhancements for Scheduled Traffic}
\acro{TDMA}{Time Division Multiple Access}
\acro{TSA}{Transmission Selection Algorithm}
\end{acronym}

\section{Introduction}
The rise of Industry 4.0 applications, e.g., intelligent factory automation, results in steadily increasing demands on communication quality.
Real-time applications, e.g., controlling the movement of a robotic arm, require guaranteed sub-microsecond end-to-end delays across multiple network nodes, so-called isochronous traffic.
As conventional Ethernet technology cannot meet these stringent \acf{QoS} requirements, the \acf{TSN} standards has been evolved to provide ultra-low latency, zero jitter, and no congestion-based packet loss.
\ac{TSN} consists of standards for time synchronization, deterministic latencies, and high reliability.
\ac{TSN} defines \ac{gPTP}, a configuration for the packet based synchronization protocol PTP that allows a nanosecond precise time synchronization of all \ac{TSN} devices.
This is the basis for any synchronous operation.
The enhancements for scheduled traffic IEEE 802.1Qbv \cites{802.1Qbv} enable the \acf{TAS} for deterministic forwarding with ultra-low delay.
It consists of eight transmission gates that control the transmission selection from the eight priority queues in the egress port.
While a gate is closed, the queue is prevented from sending, until the gate opens. 
The so-called \acf{GCL} defines the timing-based state of the gates.
By closing all gates but one, the \ac{TAS} can provide exclusive link access for a single priority omitting any delay from an already started transmission of a lower priority packet.
By carefully planning each transmission time at the Talkers and forwarding delay at the switches, \ac{TSN} implements a per-stream scheduled traffic that omits any congestion based delay.
The calculation of this \ac{TAS} schedule in combination with the precise sending time of each stream at the end stations is based on different information, such as the \ac{QoS} requirements, network topology, and traffic description.
The Industrial Internet Consortium (IIC) published the traffic types of Industrial Automation and Control Systems and maps them to the TSN technologies.
The IIC classifies critical applications into 3 classes: Isochronous, Cyclic, and Event-based traffic.
This paper focuses on the Isochronous traffic.
The calculated schedule depends on the correct arrival time of frames at bridges.
Missing, additional, early, or late frames might lead to too long end-to-end delays or packet loss, either for the end stations that deviated from their planned sending behavior or even for other end stations that strictly adhered to their planned sending behavior.
The effects of misconfigured or malfunctioning end stations for network-wide \ac{TAS} schedules have not yet been investigated.
The lack of knowledge about the fragility of \acl{PSST} in \ac{TSN} can lead to essential security measures being removed as a cost-cutting measure, resulting in fatal system failures.
The paper states four main contributions.
First, it defines the general problem statement of \ac{PSST} with misconfigured or malfunctioning end stations.
Therefore, it presents a visualization methodology for \ac{TAS} schedules over time.
It explains the effects of faulty frames, i.e., frames that are missing, additional, early, or late, and visualizes the consequences for the \ac{TAS} schedule.
Second, it quantifies the probability of errors based on faulty frames for network-wide \ac{TAS} schedules via simulation.
Finally, it demonstrates and simulates solution approaches that illustrate the necessity of time-based \ac{PSFP} to resolve the fragility issue of \ac{TAS}.

\par \medskip 

The paper is structured as follows.
We review related work in \sect{related_work} and introduce background information on \ac{TSN} in \sect{tsn}.
\sect{problem_statement} explains the problem statement and presents our novel visualization methodology for \ac{TAS} schedules.
We explain and visualize the effects of faulty frames in \sect{small-examples} and present preliminary simulation results in \sect{simulation}.
Finally, we conclude the paper in \sect{conclusion}.

\section{Related Work} 
\label{sec:related_work}
The \ac{TSN} standard "IEEE Std. 802.1Qbv Enhancements for scheduled traffic" was first published in 2018.
In the last 6 years, a lot of researchers published algorithms to create TSN schedules \cite{StOs23}.
Some of them implement reliability features within their algorithms based on specific fault models \cite{HuZh22, CrOl21, FeCa21, AtHa20, RePo20, GaZa17, FeDe22, DoDe19, PaHu21, AtHa18, ZhSa21, SyAy22, ZhSa21b, LiCh21, SyAy21b}.
Those reliability features are designed for specific faults and can be structured according to their countermeasures.
Dobirn et al. \cite{DoDe19} as well as Park et al. \cite{PaHu21} build their schedule to enable re-transmissions of missing frames within the frames deadline. Therefore, they schedule enough time between frames, such that re-transmissions do not interfere with other streams.
A similar approach that copes with frame loss is by Feng et al. \cite{FeCa21, FeDe22} and Reusch et al. \cite{RePo20}.
They send redundant copies of frames over the same path to ensure that the stream data is received even if a single frame is missing.
Zhou et al. \cite{ZhSa21} step further by adding a routing mechanism, such that the copies are no longer routed over the same path based on the probability of a frame loss, which does not necessarily result in disjoint paths.
In contrast, Atallah et al. \cite{AtHa18, AtHa20}, Gavrilut et al. \cite{GaZa17}, Syed et al. \cite{SyAy21, SyAy21b}, and Huang et al. \cite{HuZh22} ensure redundant copies that are sent via disjoint paths.
Only two of those used the IEEE Std. 802.1CB Frame Replication and Elimination for Reliability \cite{802.1CB} for this approach \cite{SyAy21b, HuZh22}.
Syed et al. \cite{SyAy23} present an encoding scheme to reduce the network capacity needed for redundancy information.
They use the XOR of two streams as a redundancy stream, which is sent via a path, disjoint of both other streams.
Therefore if one link fails, the information from the remaining two streams can be used to reconstruct the missing stream.
Another approach is fast rescheduling and rerouting after a link failure, which is implemented by Pozo et al. \cite{PoRo18}.

Crae et al. defined the Frame and Flow Isolation \cite{CrOl16}, which is implemented by some scheduling algorithms \cite{VlHa20} (Frame Isolation \cite{ZhSa21}).

Craciunas et al. introduce a scheduling that is robust against clock drifts \cite{CrOl21}.
Therefore, small sending time variations would not lead to interference between frames.

Zhou et al. implement a hardware selection that selects the \ac{TSN} hardware based on its costs, tests, and the reliability requirements of the streams \cite{ZhSa21b}.

5G ACIA white paper lists typical use cases for TSN in factory environments as well as requirements of industrial communication. The white paper states two kinds of scheduled traffic called Isochronous and Cyclic-Synchronous. For both scheduled traffic types, IEEE802.1 Qbv and Qci are mentioned as mandatory without any  reasoning.
Although Qci is mentioned as mandatory, according to the IEC/IEEE 60802, it is currently an optional feature for TSN Bridges.

\section{\acf{TSN}}
\label{sec:tsn}

We first give an overview of \acf{TSN} and explain the \ac{TAS}.
Afterwards we focus on \acf{PSST} and the \ac{PSFP}.

\subsection{Overview}

\acl{TSN} is an enhancement of Ethernet and is currently under standardization by the IEEE 802.1 \ac{TSN} task group.
\ac{TSN} specifies shaping and policing mechanisms that enable data transmission with \ac{QoS} requirements. 
Examples for \ac{QoS} requirements are bounded jitter, zero congestion-based packet loss, and sub-microsecond end-to-end delays.
End stations in a \ac{TSN} network that transmit data are called talkers, and end stations that receive data are called listeners.
Talkers and listeners are connected through possibly multiple \ac{TSN} bridges that apply shaping and policing mechanisms to provide \ac{QoS}.
Examples for shaping and policing mechanisms are \acf{CBS}~\cite{802.1Qav}, \acf{PSFP}~\cite{802.1Qch}, \acf{TAS}~\cite{802.1Qbv}, and \acf{ATS}~\cite{802.1Qcr}.

\par \medskip 

End stations signal their \ac{QoS} requirements, e.g., maximum acceptable latency, to the network before they start data transmission.
Further, they describe their data transmission in the form of an interval, also called a period, the maximum number of frames that are transmitted during the interval, and the maximum frame size.
This allows the calculation of an upper bound on the required datarate.
A logical data transmission between a talker and (multiple) listeners is called a \ac{TSN} stream.
TSN streams are admitted to the network if the \ac{QoS} demands of both the talker and listeners can be met.
This admission control decision is either done in a distributed manner or by a central configuration entity, the so-called \ac{CNC}.
Finally, bridges on the path between the talker and listeners are configured such that the previously signaled \ac{QoS} demands are fulfilled.

\subsection{\acf{TAS}}

The policing and shaping mechanisms can provide different levels of \ac{QoS} guarantees.
For example, the \acf{CBS} provides a guaranteed bandwidth share but cannot guarantee sub-microsecond end-to-end delays.
IEEE 802.1Qbv \cite{802.1Qbv} specifies the enhancements for schedule traffic, often referred to as the \acf{TAS}.
\ac{TAS} enables sub-microsecond end-to-end delays for time-critical real-time communication in \ac{TSN} networks.
It is based on a gating mechanism that introduces time slots in which frames are eligible for transmission, similar to \ac{TDMA} \cite{TDMA}.
\fig{Enhancemnets_for_scheduled_traffic.png} illustrates the structure of the Queuing mechanism in a \ac{TSN} compliant bridge with \ac{TAS}.

\figeps[\linewidth]{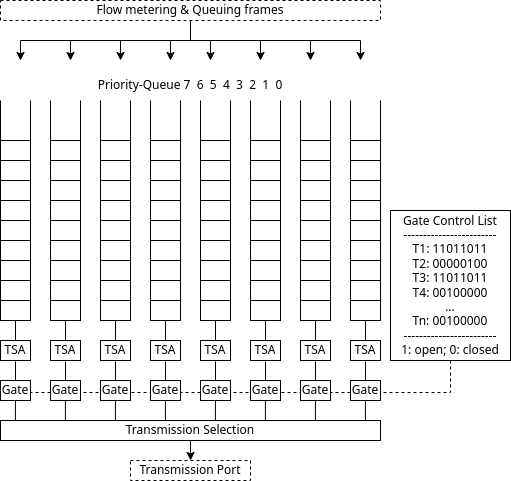}{Priority queuing and gate mechanism of the IEEE802.1Qbv Enhancements for scheduled traffic.}

\ac{TSN} bridges support up to eight priority queues per egress port.
Frames of a TSN stream are enqueued in a specific queue based on the \ac{PCP} value of its VLAN tag.
A priority queue is followed by a \ac{TSA}, e.g., \ac{CBS}, and a transmission gate with two possible states: open and close.
Frames are only eligible for transmission if their corresponding gate is open.
The \ac{GCL} controls a periodic state change of all gates, called schedule.
A \ac{GCL} entry maps a relative time interval $[T_i, T_{i+1}]$ to a bit vector, where each bit corresponds to a single transmission gate.
If the bit is set, the gate is open. 
Otherwise, it is closed.
The schedule repeats periodically for an indefinite number of times.
The length of a single schedule is called the hyperperiod, which is typically the least common multiple of all stream periods. 
If multiple queues are eligible for transmission, i.e., several transmission gates are open, the queue with the highest priority is used. 

\par \medskip 

Prioritizing or delaying streams can be done by closing the gates of other priorities at the right time. 
When multiple bridge schedules are planned together, sub-microsecond end-to-end delays can be achieved.

\subsection{Per-Stream Scheduled Traffic (PSST)}
To support streams with sub-microsecond end-to-end delays, frames need to arrive at the bridge in time, i.e., when the corresponding transmission gate is open. 
Otherwise, the frame is received while the gate is closed.
This introduces additional queuing delay, which may violate the stream's deadline.
Consequently, both the end stations and the network require precise time synchronization, e.g., with \ac{gPTP} IEEE Std. 802.1AS \cite{802.1AS}.
Additionally, the sending time at the talker and the deployed \ac{TAS} schedule needs to be compatible.
When the talker signals its \ac{QoS} requirements, it additionally sends its earliest and latest possible sending time, called transmit offset (ETO and LTO), as well as its synchronization jitter to the \ac{CNC}.
Based on these information, the \ac{CNC} calculates the sending offset for the talker and an appropriate \ac{GCL} for all bridges on the path, such that the previously signaled \ac{QoS} requirements are met.
The resulting schedule may or may not include queuing at some bridges.
Section \ref{sec:problem_statement} illustrates the consequences of different schedule decisions.

Calculating such a schedule is NP-complete and can be done by suitable algorithms \cite{StOs23}.
The combination of sending offsets at talkers and \ac{TAS} within the network is called \acf{PSST} \cite{802.1Qbv}.

\subsection{\acf{PSFP}}
\acf{PSFP} is defined with the IEEE Std. 802.1Qci in 2017 to restrict the enqueuing of unwanted packets into the priority queues at the egress ports.
Therefore it has the capability to drop packets or change the priority queue based on a combination of the time, the frame size, the rate and the burstiness of a stream.
To accomplish that, \ac{PSFP} consists of three layered components, the Stream Filters, the Stream Gates, and the Flow Meters.
\ac{PSFP} matches each arriving packet to the first Stream Filter instance where both the configured IEEE802.1CB Stream Handle and the vlan priority matches.
The Stream Filter allows a wildcard entry for both the Stream Handle and the vlan priority.
The matched Stream Filter instance may filter the frame based on the SDU Size, updates some counters, and defines the corresponding Stream Gate instance and the optional Flow Meter instance.
\figeps[\linewidth]{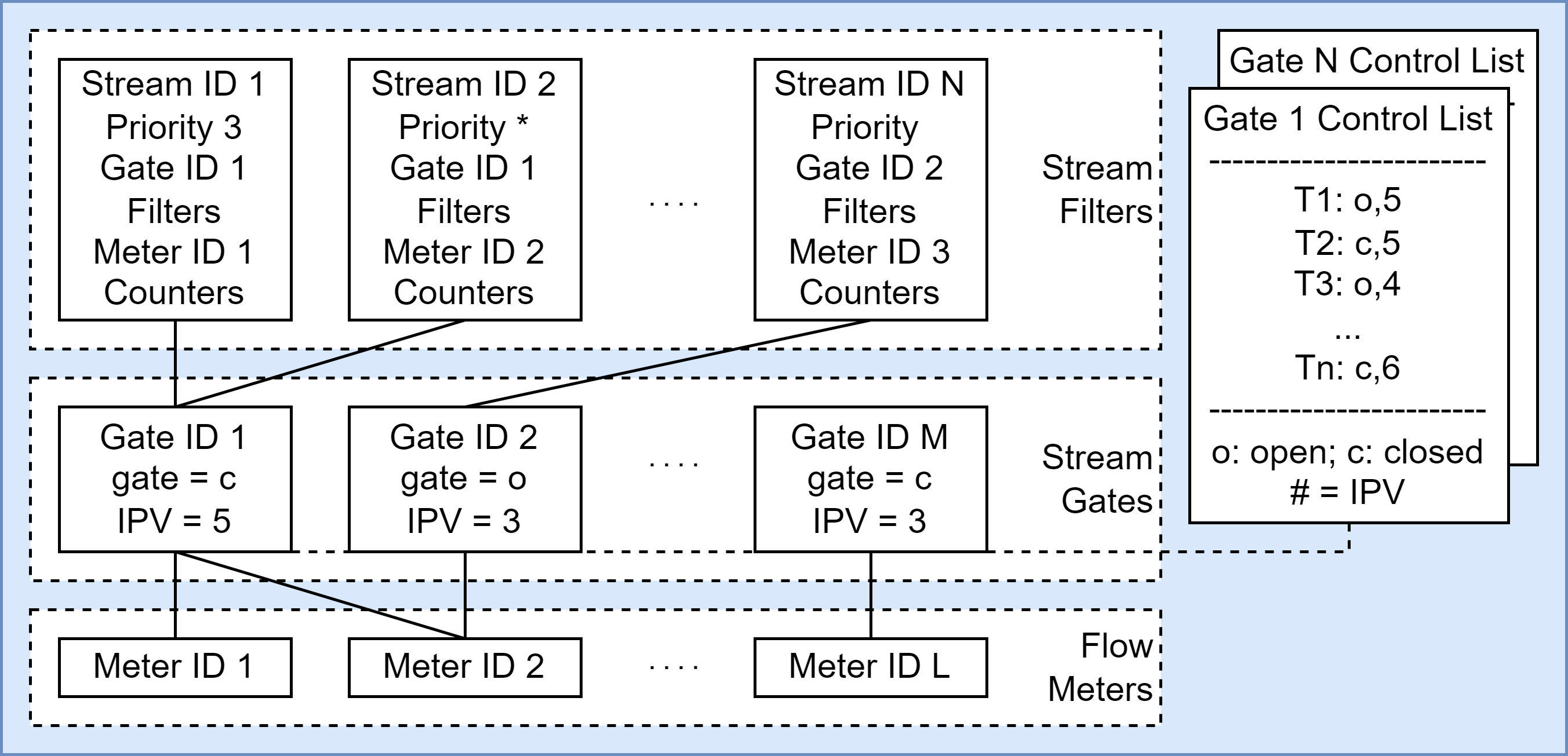}{Per-Stream Filtering and Policing}
The Stream Gate instance consists of a gate that can be opened and closed and assign an \ac{IPV} based on a timing-based gate control list.
When the packet arrives while the gate is in the closed state, the packet is dropped.
Otherwise, the packet passes the gate and might change the switch's internal priority for that frame, based on the \ac{IPV} value.
In the following, we refer to this time-based gating mechanism as time-based PSFP.

The optional Flow Meter is a two rate three color token bucket meter as defined in MEF 10.3 \cite{MEF10.3}.
The token bucket is configured using the committed information rate, committed burst size, excess information rate, and excess burst size as parameters.
This mechanism enables to filter the frames based on their data-rate.
In the following, we refer to this as rate-based PSFP.

The combination of those 3 layers enables a filtering of TSN packets based on their arrival time and data-rate.

\section{Problem Statement}
\label{sec:problem_statement}

\ac{PSST} combines precise sending times at talkers (transmission offsets) and valid \acp{GCL} in the network.
Misconfigured or malfunctioning end stations can disrupt the deployed network schedule by deviating from the planned sending behavior.
We categorize frames that deviate from their planned arrival time, so-called faulty frames, into the following groups: missing frames, additional frames, early frames, and late frames.
A missing frame is a frame that is scheduled to arrive at a specific time at a bridge but is never received by the bridge. 
The reason for missing frame might be a failed link, packet corruption (CRC failure), a broken end station, or an overloaded network component.
An additional frame is a frame received by a bridge without being planned, which can be caused by a malfunctioning or misconfigured end station.
Further, frames might arrive too early or too late at a bridge.
Reasons for too early and too late frames are misconfigured end stations, synchronization errors, and non-deterministic queuing delay in a previous network component.

\par \medskip 

Faulty frames can cause a part of the network to no longer operate as planned or cause a global network failure.
We explain and visualize the consequences of missing, additional, early, and late frames for a network-wide schedule in \sect{small-examples}.
\section{Effects of Faulty Frames in Small Examples}
\label{sec:small-examples}

In this section, we first introduce a method to visualize the behavior of an egress queue with TAS and periodic traffic. Then, we illustrate on a single link by minimal examples how faulty frames can delay later frames, possibly by significantly long time, and cause a persistent or even continuously increasing queue.

\figeps[\linewidth]{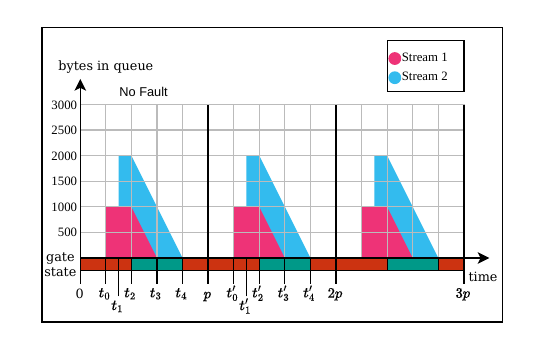}{Queue occupancy over time on an egress queue in the presence of TAS and periodic traffic w/o faulty frames. Sudden increases are frame arrivals, linear decreases indicate frame transmissions. Gate state: red means closed gate, green means open gate.}

\subsection{Visualization of Queue Behaviour with TAS and Periodic Traffic}
In \fig{no_fault} the occupancy of an egress queue is visualized over time. Frames of different streams are shown with their respective size in different colors on top of each other. In the example, the two frames are 1000 and 500 bytes large. Frames arrive periodically and instantaneously, which is indicated by a sudden increase of the queue occupancy. In contrast, we illustrate frame transmissions by linearly decreasing queue occupancy over time; the slope depends on the transmission speed\footnote{Instantaneous arrivals and continuous departures are inconsistent, but simplify the reading of the figures. When arrivals are also depicted continuous, simultaneous frame arrivals lead to superposed slopes, which looks more complex but does not add any value to the  discussion.}. In the examples, all frames have the same period, and the beginning of a period is marked by a vertical line. In case of different periods, it is more useful to mark the start of a hyperperiod. The state of the periodic TAS gate is indicated on the x-axis. The gate is closed during red intervals, and it is open during green intervals. As a consequence, frames are queued during red intervals and transmitted during green intervals.

We briefly interpret the behavior illustrated in \fig{no_fault}. At times $t_0$ and $t_1$ the magenta and blue frames arrive with 1000 and 500 bytes, respectively. This can be seen from the sudden increases of the queue occupancy at $t_0$ and $t_1$ by these values. They are queued due to the closed gate. The gate opens at $t_2$. This starts the transmission of the magenta frame, which is indicated by the continuously decreasing queue occupancy. After its completion at $t_3$, the blue frame is transmitted until $t_4$. The described behavior repeats due to the periodicity of the streams.

\figeps[\linewidth]{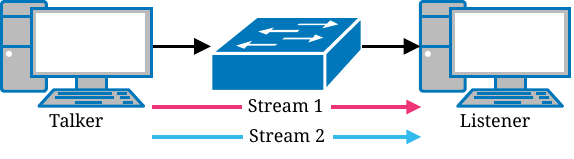}{When a single link is considered, multiple talkers send traffic over a single bridge to a listener. The queuing on the link from the bridge to the listener is considered.}

\subsection{Effects of Faulty Frames on a Single Link}
In the following, we illustrate the effects of faulty frames on a single link. The experiment setup is depicted in \fig{Network_Talker-Listener}. Possibly, multiple talkers send a stream to one listener via a single bridge. We consider the buffer occupancy on the link from the bridge to the listener. Periodic traffic is expected to arrive according to a schedule and sent within time slots controlled by the TAS. When the remaining time of the time slot is not large enough to accommodate the entire frame, the transmission of the frame is deferred.

We consider the impact of faulty frames, i.e., additional, late, early, or missing frames on potential delay of succeeding frames. Moreover, we show how unequal frame sizes can greatly increase that impact. 

\figeps[\linewidth]{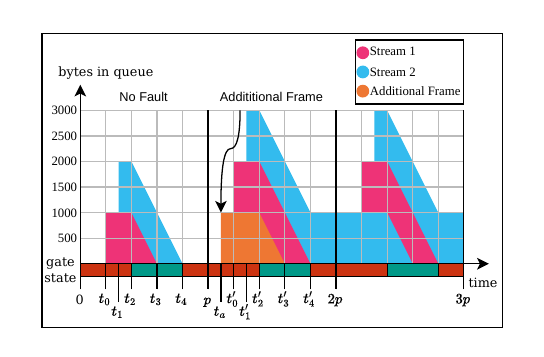}{
In the second period an additional orange frame arrives before the magenta frame. As a consequence, the orange frame takes the time slot in that period and the magenta frame is significantly delayed. The time slot taken by the additional frame is missing in the future so that the queue will contain at least one frame and cause delay for all future frames.}

\subsubsection{Impact of an Additional Frame}
\label{sec:additionalFrame}
\fig{additional} visualizes the impact of an additional frame. A magenta frame and a blue frame are scheduled for every period. They arrive in time before their respective transmission slots and are queued for a short time. The gate opens slightly after their arrival so that the frames can be sent without significant delay. However, in the second period, an additional orange frame arrives at a time $t_a$ before the arrival of the magenta frame. Therefore, the orange frame is transmitted when the gate opens, and the magenta frame remains queued until the gate re-opens again. The magenta frame then takes the transmission slot of the blue frame. When the blue frame arrives, it is queued until the gate re-opens next time, which is the transmission slot of the magenta frame. Thus, both the magenta and the blue frame miss their slots and are sent in transmission slot of other frames. A prerequisite for the latter is that there are at least two frames per period on the link. 

For the following periods holds that whenever a frame arrives, there is already another frame in the queue so that the old frame is sent while the new frame must wait again. 
This can also be viewed as follows. Normally, there are as many transmission slots as arriving frames. In case of additional frames, that number of frames is persistently buffered in the queue. As a consequence, all future frames will suffer from that queuing delay.

\subsubsection{Impact of a Late Frame}
\label{sec:lateFrame}
\fig{late-2-frames} visualizes the impact of a late frame. Here, a magenta and blue frame are scheduled for every period and normally arrive  before two distinct transmission slots. Only in the second period, the magenta frame is late. As a consequence, the magenta frame cannot be sent within its time slot because the remaining time is not sufficient. Therefore, the frame is queued until the gate re-opens again. Thereby, the frame is delayed and takes the time slot of the blue frame, which misses its time slot. 

In the next period, the effect of the delayed frame is like the effect of an additional frame: it takes the time slot of that period so that the frame arriving in that period is also significantly delayed, again. This can also be viewed as follows. If a frame is delayed and the time slot for that frame cannot be used, this time slot is missing in the future so that at least one frame remains in the queue. As a consequence, all future frames will suffer from that queuing delay.

\figeps[\linewidth]{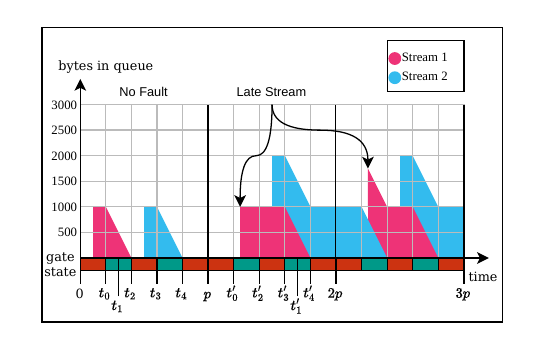}{
The magenta stream is delayed so that all its frames miss their time slots. This effects that magenta and blue frames are significantly delayed and arrive at the next link in a wrong time slot.}

\subsubsection{Impact of a Delayed Stream}
\label{sec:lateStream}
A stream is delayed if all its frames arrive late. \fig{late-2-frames} visualizes the impact of a delayed stream. Here, a magenta and a blue frame are scheduled for every period, with distinct time slots that succeed their planned arrivals. However, all magenta frames arrive slightly too late so that they miss their time slot. As a consequence, they take the time slots of the blue frames that are also delayed for the time slots of the later magenta frames. Below the line, we observe that magenta and blue frames are transmitted in different time slots. That means, all frames are delayed, and they arrive at the next link in a wrong time slot.

\figeps[\linewidth]{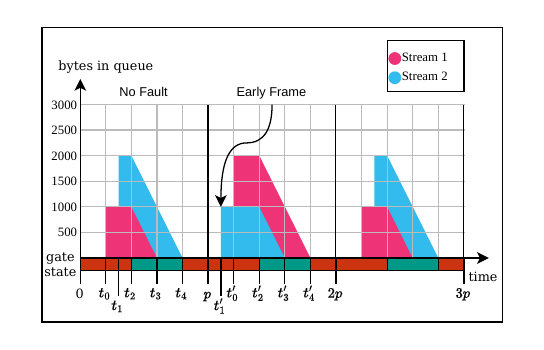}{A magenta frame usually arrives before a blue frame and is transmitted first when the gate opens. However, in the second period the blue frame arrives early and is transmitted first. Thereby, the magenta frame is sent slightly delayed. This may cause problems of a late frame on the next link (cf. \fig{late-2-frames}).}

\subsubsection{Impact of an Early Frame}
\label{sec:earlyFrame}
\fig{early} visualizes the impact of an early frame. Here, a magenta frame is scheduled and arrives before a blue frame so that it is transmitted first in the subsequent time slot. However, in the second period, the blue frame is early and therefore transmitted first. As a consequence, the magenta frame is transmitted afterwards so that it is slightly late. As the magenta frame will arrive slightly delayed at the next link, it may cause the problems of a late frame on that link (cf. \sect{lateFrame}). 

\figeps[\linewidth]{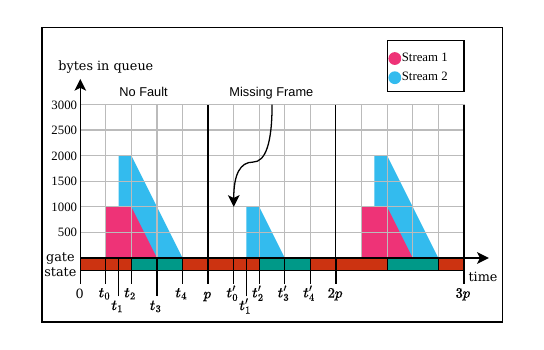}{
In the second period the magenta frame is missing so that the blue frame is transmitted as soon as the gate opens. Thereby, the blue frame is sent slightly early. This may cause problems of an early frame on the next link (cf. \fig{early}).}

\subsubsection{Impact of a Missing Frame}
\label{missingFrame}
\fig{missing} visualizes the impact of a missing frame. Here, a magenta frame is scheduled and arrives before a blue frame so that it is transmitted first in the subsequent time slot. However, in the second period the magenta frame is missing, and therefore the blue frame is transmitted as soon as the gates opens. As a consequence, the blue frame is transmitted earlier than planned. As the blue frame will arrive slightly early at the next link, it may cause the problems of an early frame on that link (cf. \sect{earlyFrame}). 

\figeps[\linewidth]{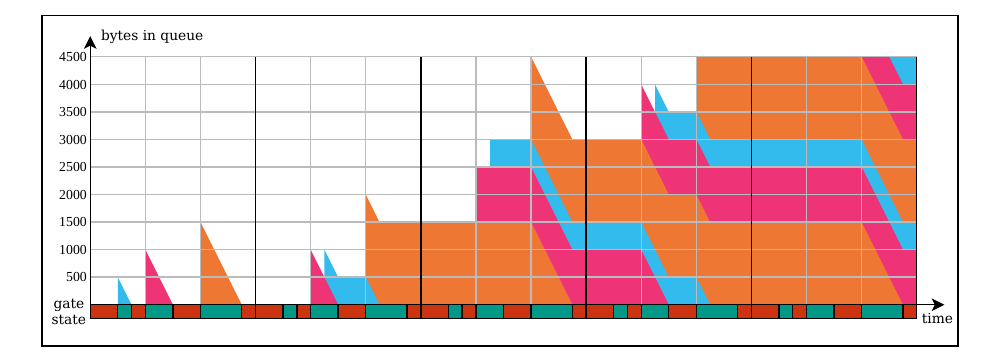}{The stream with small frame sizes is delayed, which changes the arrival order of the frame sizes relative to the sizes of the time slots. This leads to continuous increase of the queue occupancy.}

\subsubsection{Continuous Increase of Queue Occupancy in the Presence of Unequal Frame Sizes}
\fig{continuousIncrease} visualizes how a significantly delayed stream can cause a continuous increase of the queue occupancy in the presence of unequal frame sizes. We assume that three streams with small, medium, and large frames arrive within a period and their time slots succeed their frame arrivals. However, the stream with the small frames is delayed such that its frames arrive between the frames with medium and large frames sizes. This can happen due to a slightly delayed stream (cf. \sect{lateStream}). We observe that when the small frame is sent in the time slot of the large frame, the succeeding large frame waits until the time slot for the next large frame. Thereby, two time slots remain unused, which increases the queue occupancy by two frames. As this process continuous, it leads to long frame delay and in the end to frame loss.

\section{Simulation and Validation}\label{sec:simulation}
After showing the possible impacts of timing deviations in theory, this section simulates a small network using the network simulator OMNeT++ and its INET framework \cite{OMNeT, INET}.
The OMNeT++ simulations consist of three parts.
First, the network topology file (Network.ned), which defines the end stations, TSN switches and the connections between them.
Second, the OMNet++ simulation initialization file (omnetpp.ini) that contains the configuration of all endstations and switches, including the GCL for all switches and the scheduled sending times for each stream.
And third, a scenario file that controls the fault generations.
To enable a stable fault generation, we had to apply a patch to the application generation, which we provide with all simulation files and instructions on GitHub.
The simulations omit the \ac{gPTP} time synchronization procedure to avoid interference between time synchronization and TSN applications, and prefill the MAC tables of all switches to prevent broadcast messages at the start of each simulation.

\subsection{Validation of the Theoretical Results}
To validate the theoretical results of Section \sect{small-examples}, all experiments were simulated with OMNeT++.
The gate state over time as well as the transmission times were compared between the event log of the simulation and the above graphs.
Both versions showed the same behavior.
The source code for those simulations, together with a manual are published on GitHub (\url{https://github.com/EpplerM/TSN-Fault-Simulation}).

\subsection{Preliminary Simulation}
\figeps[\linewidth]{affectedNetwork.pdf}{Simulated network for Validation}
Fig.~\ref{fig:affectedNetwork.pdf} shows a simple network topology consisting of 5 bridges and 5 end-stations, connected with gigabit Ethernet links.
Each bridge is connected to one end-station and two or three bridges.
The end-stations send and receive scheduled streams, and the bridges use a TAS to forward them according to a predefined schedule.
Table~\ref{tab:streamsPropagationExample} shows the deployed streams as well as their scheduled path.
\begin{tab}{streamsPropagationExample}{Streams of the propagation example.}
    \begin{tabular}{c|c|c|c}
        Stream & Talker & Path & Listener \\\hline\hline
        A & ES 1 & Bridge 1, Bridge 2, Bridge 3 & ES 3\\\hline
        B & ES 2 & Bridge 2, Bridge 3, Bridge 4 & ES 4\\\hline
        C & ES 3 & Bridge 3, Bridge 4, Bridge 1 & ES 1\\\hline
        D & ES 4 & Bridge 4, Bridge 1, Bridge 2 & ES 2\\\hline
        E & ES 3 & Bridge 3, Bridge 4, Bridge 5 & ES 5\\\hline
        F & ES 4 & Bridge 4, Bridge 5, Bridge 2 & ES 2\\\hline
        G & ES 5 & Bridge 5, Bridge 2, Bridge 1 & ES 1\\
    \end{tabular}
\end{tab}
The streams are sent periodically with a period length of 60 $\mu s$ and a frame size between 250 and 1250 Bytes.
\figeps[\linewidth]{Latency_broken_ring.png}{End-to-End Latency of the applications A to G within the Network. After the fault at 10 ms, the latency of all frames increases over time.}
While all streams are sent as scheduled, all frames have a constant latency of less than 100 $\mu s$.
\subsection{Preliminary Results}
A fault is introduced after 10 ms, by sending a single frame from stream A 10 $\mu s$ later than scheduled.
As a result, the frame remains in the queue in the same way as the example in Section \ref{sec:lateFrame}.
Fig.~\ref{fig:Latency_broken_ring.png} shows an increasing delay for all streams, after the fault occurred.
After less than 500 ms, all frames experience a delay of more than 10 times their regular delay.
The simulation shows, that a system using PSST with the TAS, a delayed frame may lead to a fatal network error.

\section{Conclusion}
\label{sec:conclusion}
\ac{PSST} in \ac{TSN} depends on the \ac{TAS} and the sending time at the talkers.
This results in a very fragile communication, as even a small problem may result in a fatal network error.

TSN scheduled traffic ist sehr fragile. Einzelnes Problem kann gesamten scheduled traffic lahmlegen.
Folge ist Frabrikausfall.
TSN scheduled traffic muss immer mit iwas abgesichert werden.

Naive ansätze wie hinzufügen von Slots für Fehler und vergrößern der Slots kann die Wahrscheinlichkeit für Fehler senken, das ursprüngliche Problem jedoch nicht lösen.

PSFP kann mit trivialem schedule die fehlerhaften frames droppen, was das Problem teilweise löst.
In Kombination mit scheduling das Frame isolation implements, this solves the fragility.

TSN scheduled traffic needs Frame isolation and PSFP to ensure a working network.

Current Hardware is not capable of reliable PSFP filtering for each stream.
This needs to be solved before TSN can be implemented in industrial environments.

In Zukunft muss methode gefunden werden um gegen Talking Idiots und Angreifer zu schützen.
Wir vermuten, dass eine Kombination aus Frame Isolation und PSFP dies schaffen könnte.

The reliability of \ac{PSST} in \ac{TSN} strongly depends on the \ac{TAS} configuration and the precise sending times of the talkers. Consequently, TSN scheduled traffic is inherently fragile -- even minor timing deviations or configuration errors can lead to severe communication failures, potentially resulting in complete factory downtime.

Therefore, TSN scheduled traffic must always be safeguarded by additional mechanisms. 
The \ac{PSFP} mechanism can mitigate part of this fragility by dropping faulty frames using a simple schedule. When combined with scheduling strategies that implement frame isolation, the overall robustness of TSN communication can be significantly improved. Thus, ensuring reliable TSN operation requires both PSFP and frame isolation.

However, most current hardware implementations are not yet capable of performing reliable PSFP filtering for every stream. This limitation must be resolved before TSN can be deployed in safety-critical industrial environments.

Future research should focus on developing methods to protect TSN networks from both unintentional misbehavior (“talking idiots”) and deliberate attacks. We hypothesize that a combination of frame isolation and PSFP could provide an effective foundation for such protection.

\AtNextBibliography{\small}
\printbibliography
\end{document}